\documentclass[11pt]{article}
\usepackage{alpha}
\usepackage{amsmath,amsfonts,epsfig,graphics}
\usepackage{amssymb}
\usepackage{amsbsy,color}
\usepackage{siunitx}
\usepackage{physics}
\usepackage{graphicx}
\usepackage{psfrag}
\usepackage{multirow}
\usepackage{booktabs}
\usepackage{url}
\usepackage{dcolumn}
\usepackage[T1]{fontenc}

\newcommand{\be}{\begin{equation}}
\newcommand{\ee}{\end{equation}}
\newcommand{\bea}{\begin{eqnarray}}
\newcommand{\eea}{\end{eqnarray}}
\newcommand{\bes}{\begin{eqnarray}}
\newcommand{\ees}{\end{eqnarray}}
\newcommand{\ba}{\begin{array}}
\newcommand{\ea}{\end{array}}

\newcommand{\Eq}[1]{Eq.~(\ref{#1})}

\newcommand{\fig}[1]{Fig.~\ref{#1}}

\newcommand{\Sect}[1]{Section~\ref{#1}}

\newcommand{\Tab}[1]{Table~\ref{#1}}

\newcommand{\Ref}[1]{Ref.~\cite{#1}}

\newcommand{\chitop}{\chi_{_\mathrm{YM}}}

\newcommand{\rmO}{\mathrm{O}}

\newcommand{\su}[1]{$\mathrm{SU}(#1)$}

\preprintno{DESY 16-131}

\pdfoutput=1   

\begin{document}
\title{The topological susceptibility in the large-$N$ limit of $\mathrm{SU}(N)$ Yang-Mills theory}

\author[pi,pi2]{Marco C\`e}
\author[desy,hu]{Miguel Garc\'ia Vera}
\author[mi1,mi2]{Leonardo Giusti}
\author[desy]{Stefan~Schaefer}

%
%
\address[pi]{%
  Scuola Normale Superiore,
  Piazza della Cavalieri 7, I-56126 Pisa, Italy
}
\address[pi2]{%
  INFN, sezione di Pisa,
  Largo B. Pontecorvo 3, I-56127 Pisa, Italy
}
\address[desy]{John von Neumann Institute for Computing (NIC), DESY, \\
    Platanenallee 6, D-15738 Zeuthen, Germany}

\address[hu]{
  Insitut f\"ur Physik, Humboldt Universit\"at zu Berlin,\\
  Newtonstr. 15, D-12489 Berlin, Germany
}

\address[mi1]{%
  Universit\`a di Milano Bicocca,
  Piazza della Scienza 3, I-20126 Milano, Italy
}
\address[mi2]{%
  INFN, sezione di Milano Bicocca,
  Piazza della Scienza 3, I-20126 Milano, Italy
}

\begin{abstract}
We compute the topological susceptibility of the $\mathrm{SU}(N)$ Yang-Mills theory in the
large-$N$ limit with a percent level accuracy. This is achieved by measuring the gradient-flow
definition of the susceptibility at three values of the lattice
spacing for $N=3,4,5,6$.  Thanks to this coverage of
parameter space, we can extrapolate the results to the large-$N$ and continuum limits with
confidence. Open boundary conditions are instrumental to make simulations feasible on the finer
lattices at the larger $N$.
\end{abstract}


\maketitle

\section{Introduction} 
The limit of large number of colors $N$ has proved to be a fruitful tool
in the study of \su{N} Yang--Mills theories~\cite{'tHooft:1973jz}.
One example is the Witten-Veneziano formula explaining the large value of
the mass of the $\eta'$ meson in the chiral limit~\cite{Witten:1979vv,Veneziano:1979ec} 
\begin{equation} 
\lim_{N \to \infty} \frac{m_{\eta'}^2 F_\pi^2}{2 N_\mathrm{f}} = \lim_{N \to \infty} \chi_{_\mathrm{YM}}
\qquad \text{with} \qquad \chi_{_\mathrm{YM}}=\int d^4x\, \langle q(x)\, q(0) \rangle_{_\mathrm{YM}} \,, 
\label{eq:wv}
\end{equation}
where $F_\pi$ is the pion decay constant, $N_\mathrm{f}$ the number of massless quark flavors, and 
${q=\frac{1}{32\pi^2} \epsilon_{\mu\nu\rho\sigma} \tr F_{\mu\nu} F_{\rho\sigma}}$ the topological
charge density. 
This formula can be given a precise meaning in quantum field theory by properly
defining the topological susceptibility $\chi$ in QCD and in the Yang--Mills
theory~\cite{Giusti:2001xh,Seiler:2001je,Giusti:2004qd,Luscher:2004fu}.
The value of $\chi_{_\mathrm{YM}}$ found in 
\su{3} Yang--Mills theory~\cite{DelDebbio:2004ns} is large enough to solve the $\mathrm{U}(1)_\mathrm{A}$
problem in QCD, a fact which makes it extremely interesting to study its value in the
large-$N$ limit.

Exploratory computations with cooling techniques at large $N$ have a long tradition on the
lattice~\cite{Lucini:2001ej,DelDebbio:2002xa,Lucini:2004yh}, with quoted errors
for the topological susceptibility at the $10\%$ level. These results, however,
reflect the short-comings of the techniques available at the time. In particular, a theoretically sound
definition of the topological susceptibility with a well-defined and universal continuum
limit had not been used. Only \Ref{Cundy:2002hv} opted for the theoretically clean but
expensive definition via the index of a chiral Dirac operator, and was therefore limited
to a very coarse lattice spacing and small statistics. 

The second problem affecting all simulations concerned with topological quantities is
the quickly freezing  topological charge as the continuum limit is approached.
At large values of $N$ this makes it exceedingly hard to perform reliable simulations at
small lattice spacings, since the number of updates needed rises dramatically with the
inverse lattice spacing \cite{DelDebbio:2002xa,Schaefer:2010hu}. This comes on top of
the increase of the cost of the updates growing with $N^3$, such that it cannot be
overcome by a brute force approach.

Taking advantage of the conceptual, algorithmic and technical developments of the
last decade, we are in the position to improve significantly over these results.
The exceptional slowing down of the topological modes can be avoided by using open
boundary conditions in time~\cite{Luscher:2011kk}. With the introduction of the
gradient flow, a theoretically clean and numerically cheap definition of the topological
charge has become available~\cite{Luscher:2010iy,Luscher:2010ik}. In the
continuum limit the corresponding topological susceptibility satisfies the singlet chiral
Ward identities when fermions are included, and is the proper quantity to be inserted
in the Witten--Veneziano formula~\cite{Ce:2015qha}. 

The aim of this Letter is to compute the topological susceptibility in the large-$N$ and
continuum limits with percent accuracy. We measure $\chi_{_\mathrm{YM}}$ for the groups
\su{4}, \su{5} and \su{6}, and combine the results with previous ones for
\su{3}~\cite{Ce:2015qha}. Since leading corrections are expected to be $\rmO(N^{-2})$,
this gives us a factor of four in their size. For each group the three lattice spacings simulated
range from $0.096$~fm to $0.065$~fm with leading $\rmO(a^2)$ discretization effects decreasing by
more than a factor $2$ in size. This coverage of parameter space allows for a robust extrapolation
of the results to the large-$N$ and continuum limits.

This Letter starts with giving the continuum definitions of the observables
in \Sect{s:2} followed by the details of the lattice setup in \Sect{s:3}. The
extrapolations to the continuum and large-$N$ limit, giving the final results, are
presented in \Sect{s:4} before some concluding remarks.

\section{Observables\label{s:2}}
The Yang--Mills gradient flow has proved to be a very versatile tool to define 
a variety of observables with a smooth continuum limit~\cite{Narayanan:2006rf,Luscher:2010iy}.
It evolves, in the continuum, the gauge field $B_\mu$ as a function 
of the flow time $t\ge 0$ solving the initial value problem~\cite{Luscher:2010iy}
\begin{equation}
\label{eq:YMgf_def}
  \partial_t B_\mu = D_\nu G_{\nu\mu}  \;, \quad \eval{B_\mu}_{t=0} = A_\mu \;,
\end{equation}
where
\begin{equation}
  G_{\mu\nu} = \partial_\mu B_\nu - \partial_\nu B_\mu - i \comm{B_\mu}{B_\nu} \;, \quad D_\mu = \partial_\mu - i \comm{B_\mu}{\cdot} \;,
\end{equation}
and thus providing a Gaussian smoothing of the gauge fields with a radius 
$\sqrt{8t}$. We are interested in the energy density $e^t$ and the topological charge
density $q^t$ at flow time $t$, which are defined as 
\begin{equation}
\label{eq:e_cont_def}
  e^t(x) = \frac{1}{2} \tr \big[ G_{\mu\nu}(x) G_{\mu\nu}(x) \big] \quad \text{and} \quad
  q^t(x) = \frac{1}{32\pi^2} \epsilon_{\mu\nu\rho\sigma} \tr\big [ G_{\mu\nu}(x) G_{\rho\sigma}(x) \big] \;.
\end{equation}
The power of the flow resides in the fact that at $t>0$ operators made up of evolved
fields, such as $e^t(x)$ and $q^t(x)$, are finite as they stand once inserted in correlation
functions, i.e. no ultraviolet renormalization is required. Moreover, short-distance singularities
cannot arise, and integrated correlators are well defined. These properties carry over to the discretized theory,
where (integrated) correlators have a finite and universal continuum limit as they stand.

Thanks to the topological nature of $q^t$, continuous deformations of the gauge field induced
by the gradient flow do not affect the cumulants of the topological charge, which, in the
continuum theory, are constant along the flow~\cite{Luscher:2010iy,Ce:2015qha}. 

\subsection{Definition of the reference scale $t_0$}
In order to relate results in theories with different $N$, we need to
define a reference scale in terms of which the observables
are expressed. While different choices are logically possible, it is
desirable to choose a quantity which is a (non-zero) constant at leading order in $1/N$,
and that can be computed with high numerical precision. We opt for generalizing $t_0$ proposed for
$N=3$ in Ref.~\cite{Luscher:2010iy} to arbitrary values of $N$, by requiring
\begin{equation}
\label{eq:t0_def}
  \eval{t^2 \ev{e^t}}_{t=t_0} = 0.1125\, (N^2-1)/N \;,
\end{equation}
such that the right hand side attains the canonical value of $0.3$ for \su{3}.
At small $t$, perturbation theory gives
\begin{equation}
  t^2 \ev{e^t} = \frac{3}{128\pi^2} \frac{N^2-1}{N} \lambda_t(q) 
  \left[ 1 + c_1 \lambda_t(q) + \order{\lambda_t^2} \right]\; , 
\end{equation}
where $\lambda_t(q)=N g^2(q)$  at the scale $q=(8t)^{-\flatfrac{1}{2}}$
is the renormalized 't~Hooft coupling, and $c_1 = \frac{1}{16\pi^2}
( \frac{11}{3}\gamma_E + \frac{52}{9} -3\ln{3})$. The sub-leading term
on the r.h.s. of Eq.~(\ref{eq:t0_def}) has been included following
the indication of the perturbative expression.

Since $\text{SU}(N)$ Yang--Mills theory is not realized in Nature, any
conversion of this result to physical units is a matter of convention.
For the sake of clarity in the presentation, however, it is useful to assign
a physical value to $t_0$, which we choose to be $\sqrt{t_0}=0.166\,\mathrm{fm}$
for all values of $N$. This is motivated by the fact that in the \su{3} theory
$\sqrt{8t_0}/r_0=0.941(7)$~\cite{Ce:2015qha}, together with a  value of the
Sommer scale $r_0=0.5\,\mathrm{fm}$\,\cite{Sommer:1993ce}.
We will use this value of $t_0$ to express the lattice sizes and lattice spacings in
physical units,
but not to convert the final results, which instead will be expressed always in units of $t_0$.

For completeness, it is useful to remember that in the \su{3} theory 
$\sqrt{t_0}\, \Lambda_{\overline{\mathrm{MS}}}=0.200(16)$~\cite{Capitani:1998mq},
where $\Lambda_{\overline{\mathrm{MS}}}$ is the lambda parameter of the theory.
It would be desirable in the future to compute this quantity at higher
$N$, and eventually take the $N\rightarrow\infty$ limit.

\section{Lattice details\label{s:3}}
The standard discretization of  $\mathrm{SU}(N)$ Yang--Mills theory on
four-dimensional lattices of size $T\times L^3$ and lattice spacing $a$ 
is used throughout this study. We use the  Wilson plaquette action 
\begin{equation}
  S_\text{W}[U] = \beta \sum_P w_P \left( 1 - \frac{1}{N} \Re\tr U_P \right) \;, \quad \beta \equiv \frac{2N}{g_0^2} = \frac{2N^2}{\lambda_0} \;,
\end{equation}
where $U_P$ is the ordered product of links around the plaquette $P$,
$\lambda_0$ is the bare 't~Hooft coupling, and $w_P=1$ everywhere except
for the space-like plaquettes on the time slices $0$ and $T-a$
where $w_P=1/2$~\cite{Wilson:1974sk}. This because we opted for
open boundary conditions in time as implemented in Ref.~\cite{Luscher:2011kk},
while spatial directions are periodic.

\begin{table}[tb]
  \centering
  \begin{tabular}{ccccccccc}
    \toprule
    {\#run} & {$N$} & {$\beta$} & $1/\lambda_0$ & {$T/a$} & {$L/a$} & {$a[\si{\femto\meter}]$} & {\#meas.}   & {\#it.} \\
    \midrule
    $A(4)_1$ & 4 & 10.92 &0.3413&  64 & 16  &   0.096 & 22k&   40 \\
    $A(4)_2$ & 4 & 11.14 &0.3481&  80 & 20  &   0.078 & 41k&   80 \\
    $A(4)_3$ & 4 & 11.35 &0.3547&  96 & 24  &   0.065 & 21k&  160 \\
    \midrule                                          
    $A(5)_1$ & 5 & 17.32 &0.3464&  64 & 16  &   0.095 & 15k&  120 \\
    $A(5)_2$ & 5 & 17.67 &0.3534&  80 & 20  &   0.077 & 27k&  240 \\
    $A(5)_3$ & 5 & 18.01 &0.3602&  96 & 24  &   0.064 & 14k&  480 \\
    \midrule                                          
    $A(6)_1$ & 6 & 25.15 &0.3493&  64 & 16  &   0.095 & 30k&  250 \\
    $A(6)_2$ & 6 & 25.68 &0.3567&  80 & 20  &   0.076 & 17k&  500 \\
    $A(6)_3$ & 6 & 26.15 &0.3632&  96 & 24  &   0.063 & 16k&  450 \\
    \bottomrule
  \end{tabular}
  \caption{Parameters of the simulation. For each of the three gauge groups \su{N} we
  give the inverse coupling $\beta$, the inverse of the 't~Hooft coupling $\lambda_0=g_0^2 N$ to four significant digits, the dimensions of the lattice, the 
  approximate lattice spacing using $\sqrt{t_0}=0.166$\,fm followed by the number of
  measurements and their separation in Cabibbo--Marinari updates of the lattice.
  \label{tab:simulations}
  }
\end{table}
The parameters of the simulation are collected in Table~\ref{tab:simulations}:
for each of the three gauge groups $\mathrm{SU}(4)$, $\mathrm{SU}(5)$ and
$\mathrm{SU}(6)$, three values of  $\beta$ are chosen such as to give
approximately the same $t_0/a^2$. Using $\sqrt{t_0}=\num{0.166}$\,fm, they
correspond to lattice spacings of approximately $\num{0.096}$, $\num{0.078}$ and
$0.065$~fm. The size of the boxes have been scaled such that
$L\approx 1.5~\mathrm{fm}$, while the temporal extent is chosen to be
$T=4L$, so that a sufficiently large bulk region with negligible boundary
effects is available for the measurements.

\subsection{Wilson flow observables}
We employ the standard discretization of the Wilson flow. It is integrated 
with the third-order Runge--Kutta integrator defined in~\cite{Luscher:2010iy}
with an integration step size such that the integration error is well below 
the statistical accuracy of the observables.
The two primary observables $e^t(x)$ and $q^t(x)$ are measured with a $0.04a^2$
resolution in the flow time $t$ and interpolated quadratically from the neighboring points
to get the observables at arbitrary values of $t$.

Following again Ref.~\cite{Luscher:2010iy}, the discretized 
$e^t(x)$ and $q^t(x)$ are defined through the standard ``clover'' field
strength tensor and immediately summed over the spatial directions
\begin{equation}
  \bar{e}^t(x_0) = \sum_{\vec{x}} e^t(\vec{x},x_0) \;, \qquad 
  \bar{q}^t(x_0) = \sum_{\vec{x}} q^t(\vec{x},x_0) \;.
\end{equation}

Because of the open boundary conditions, time translation invariance is broken
and some care must be taken when averaging over the $x_0$ coordinate. A 
plateau range needs to be determined, where boundary effects can be neglected.
To this end, for each observable we first
perform a fit to the symmetrized data using the contribution of one excited state 
$f(x_0)=A+B e^{-x_0 m}$ in a region where this ansatz describes the data well. 

With this result, we determine the minimal distance of the plateau fit from the
boundary requiring that $|f(d)-A|<\sigma/4$, with $\sigma$ being the average
error of the measurement for $x_0>d$. Using this criterion, the choice of
$d=9.5\sqrt{t_0}$ guarantees that boundary effects in  $\bar{e}^t(x_0)$ at
$t=t_0$ are negligible with our statistics, and therefore we define
\be
\ev{e ^t}  = \frac{a^4}{(T-2d)\,L^3}
\sum_{x_0=d}^{T-a-d} \ev{\bar{e}^t(x_0)}\; . 
\ee

\subsection{Topological susceptibility}

For the topological susceptibility, we use the approach of Ref.~\cite{Bruno:2014ova}. 
The topological charge correlator is to be averaged over the bulk region given
by a minimal distance $d$ from the boundaries
\begin{equation}
  \bar{C}^t(\Delta) = \frac{a^4}{(T-2d - \Delta)L^3}
                      \sum_{x_0=d}^{T-a-d-\Delta} \ev{\bar{q}^t(x_0) \bar{q}^t(x_0 + \Delta)} \,.
\end{equation}
Again we determine $d$ such that for all values of $\Delta$ boundary effects are negligible.  Using
the same strategy as for the energy density above,
$d=7.5\sqrt{t_0}$ turns out to be a conservative choice for all ensembles. 

An estimator of the  topological susceptibility is then obtained by truncating the sum over 
$\Delta$ with a cut-off $r$
\begin{equation}
\label{eq:chi_corr_def}
  \chitop^{t,\text{corr}}(r) = \bar{C}^t(0)+ 2 \sum_{\Delta=a}^r  \bar{C}^t(\Delta) \;,
\end{equation}
where $r$ has to be chosen such that the contribution of the neglected tail is insignificant compared to the 
statistical accuracy of the result. Such an $r$ can always be found, because the correlator
converges exponentially to zero for large separations $\Delta$. Unfortunately, the
combination of the smoothing by the gradient flow and the numerical errors
obscure this behavior in the actual data.

Due to the smoothing, the correlation function $\bar{C}^t(\Delta)$ is positive
for small values of $\Delta$ and would be expected to turn negative before
exponentially converging to zero for $\Delta \gg \sqrt{8 t_0}$. We cannot
resolve this latter feature due to the numerical uncertainties of our data,
the correlator being zero within errors typically from $\Delta=5\sqrt{t_0}$
on. 

In order to get a better handle on the contribution of the tail, we use
high precision data in \su{3}~\cite{Vera:2016xpp}. Assuming that the
relative contribution of the tail does not change drastically with $N$,
and given the accuracy of our data, cutting the summation over $\Delta$ at
$r=7\sqrt{t_0}$ is a conservative choice, leading to a negligible systematic
error. 

\subsection{Finite volume}
By their nature, lattice simulations are done in a finite volume, which
can distort the results. For a large enough lattice dimension, these systematic
effects are exponentially suppressed, but we need to verify that they are
negligible given the target accuracy.

The lattices employed in this study are slightly larger than the ones used for
\su{3} in \Ref{Ce:2015qha}. Despite significantly smaller statistical
errors, no significant finite size effects could be detected in the \su{3} study. 
In order to avoid relying only on the independence of these finite volume effects on
$N$, we also generated lattices with $L=1.1$\,fm and $2.3$\,fm for \su{4} and
\su{5} at the smallest values of $\beta$. These lattices bracket the $L=1.5$~fm
used in our analysis. With a numerical accuracy matching our target, 
no significant differences between the three sizes
are found, such that we conclude that also this systematics is under control.

\subsection{Autocorrelations}
Simulations like the one presented here are known to be challenging due to a
rapid rise of the autocorrelation times $\tau_\mathrm{int}$, in particular of
topological observables. Numerical evidence suggests they increase with a
very high power or even exponentially with $1/a$ and $N$ when periodic boundary
conditions are implemented~\cite{DelDebbio:2002xa}. 

In our study, the gauge field is updated with the Cabibbo--Marinari
scheme~\cite{Cabibbo:1982zn}: one update consists of a heat bath sweep of the
full lattice followed by $n_\mathrm{ov}\propto a^{-1}$ overrelaxation sweeps.
Both the heat bath and the overrelaxation sweeps update all the
$\flatfrac{N(N-1)}{2}$ $\mathrm{SU}(2)$ subgroups of a given $\mathrm{SU}(N)$
link.
The number of these updates between measurements is given in
\Tab{tab:simulations} and chosen such that for all our observables 
autocorrelations are hardly detectable. We take them into
account in our analysis.

To study the effect of the open boundaries, we have computed $\tau_\text{int}$
for the coarser lattices $A(4)_1$, $A(5)_1$ and $A(6)_1$ with dedicated runs in
the presence of periodic and open boundaries, putting fewer updates between
measurements for increased sensitivity.

In units of updates with $n_\mathrm{ov}=\flatfrac{L}{(2a)}$, $\tau_\text{int}$
of $\chitop$ for the periodic lattices is $16(2)$, $54(6)$ and
$187(19)$ for $N=4$, $5$ and $6$, respectively. With open
boundaries the corresponding values are $12(1)$, $46(6)$ and $111(10)$.
For all values of $N$ we observe a reduction in  $\tau_\mathrm{int}$ for
open compared to periodic boundary conditions. It is most significant at $N=6$
and hardly statistically significant for the other values of $N$.
  
Going to finer lattices, this advantage is expected to be more pronounced. The
exponential scaling observed in Ref.~\cite{DelDebbio:2002xa} with periodic
boundary conditions would suggest a value of $\tau_\mathrm{int}$  one or two orders of
magnitude larger than the one we observe with open boundaries. Therefore the finer lattice
spacings would not have been feasible with our computer resources.

\section{Results\label{s:4}}
\begin{figure}[t!]
\begin{center}
  \includegraphics[width=0.49\textwidth]{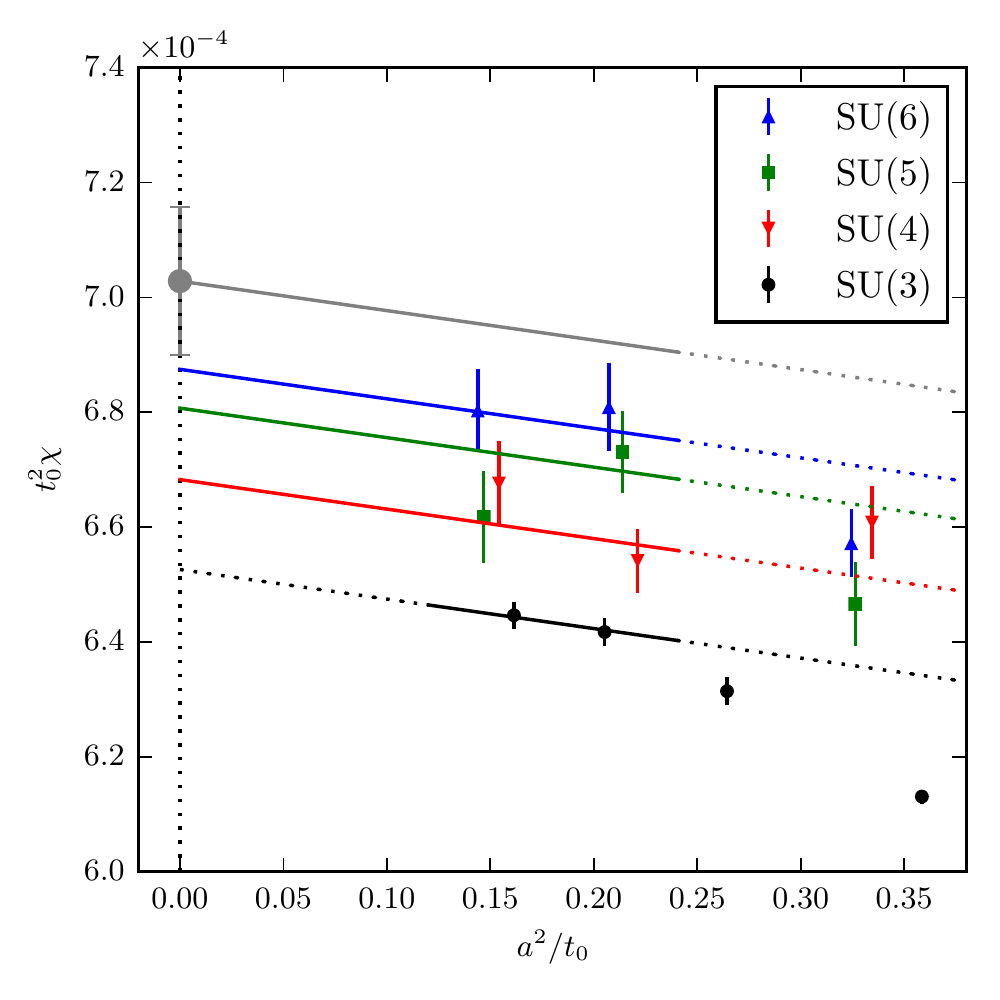}
  \includegraphics[width=0.49\textwidth]{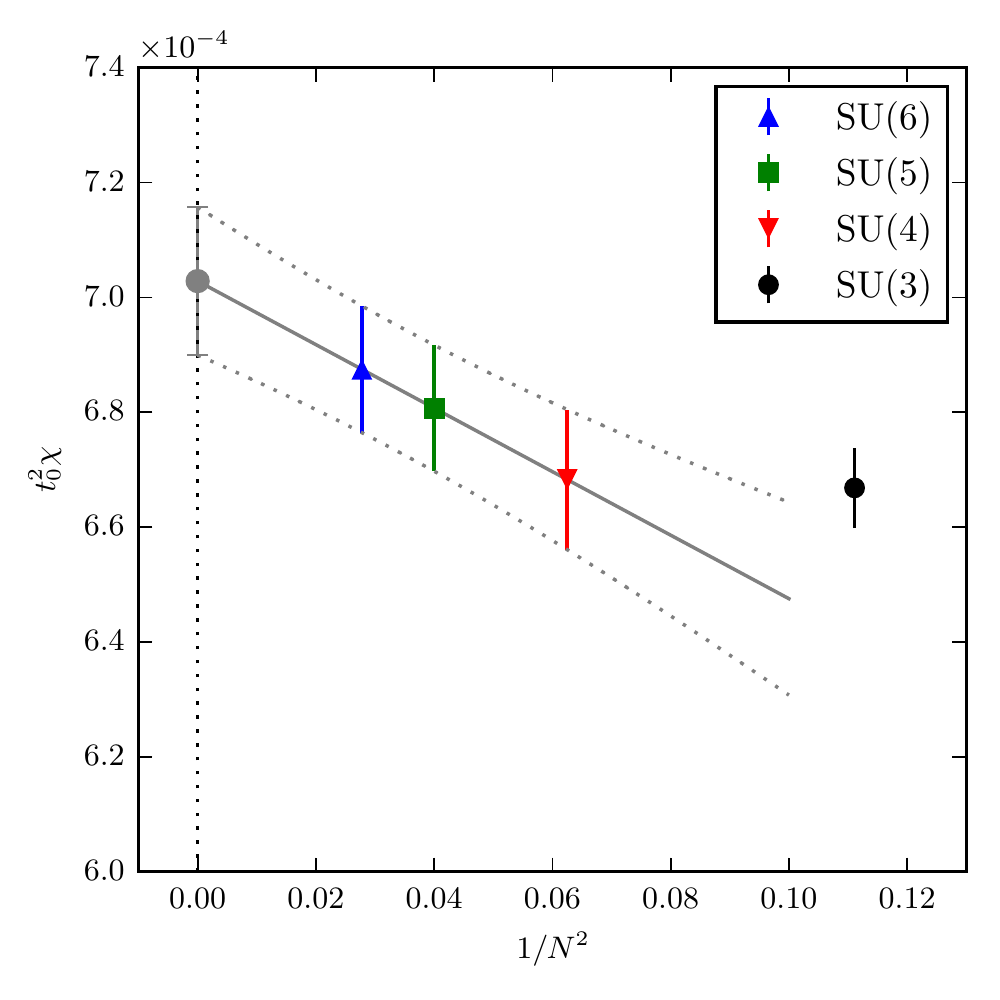}
\end{center}
\caption{\label{fig:1} Left: Combined continuum and large-$N$ extrapolation of the topological susceptibility
by fitting the data in the region indicated by the solid lines to Eq.~(\ref{e:gf}). 
Right: Same data as in left plot as a function of $1/N^2$. The data points indicate the continuum result of the 
global fit for $N>3$. For $N=3$ the data is taken from Ref.~\cite{Ce:2015qha}.}
\end{figure}

The results for the observables for the three gauge groups  are listed in
\Tab{tab:chir2}. At finite lattice spacing, we reach accuracies on
the percent level for $\chitop$ and below the permille level for $t_0$. The
values for the dimensionless product $t_0^2\chitop$  are displayed in the left plot
of \fig{fig:1}, where we also add the \su{3} results from \Ref{Ce:2015qha}. It is
clear that, both, the effects of finite $N$ and finite cut-off $a$ are roughly
at the level of our statistical errors.

\begin{table}[h]
\centering
\begin{tabular}{ccccc}
\toprule
ensemble  & $t_0/a^2$       & $10^4 \, t_0^2 \chitop$ & $ t_0^{1/2} \chitop^{1/4}$ \\\midrule
$A(4)_1$  & $2.9900(7)$   &  $6.61(6)$       & $0.1603(4)$               \\
$A(4)_2$  & $4.5207(8)$   &  $6.54(5)$       & $0.1599(3)$               \\
$A(4)_3$  & $6.4849(16)$  &  $6.68(7)$       & $0.1607(4)$               \\\midrule
$A(5)_1$  & $3.0636(7)$   &  $6.47(7)$       & $0.1595(4)$               \\
$A(5)_2$  & $4.6751(8)$   &  $6.73(7)$       & $0.1611(4)$               \\
$A(5)_3$  & $6.8151(17)$  &  $6.62(8)$       & $0.1604(5)$               \\\midrule
$A(6)_1$  & $3.0824(4)$   &  $6.57(6)$       & $0.1601(4)$               \\
$A(6)_2$  & $4.8239(9)$   &  $6.81(8)$       & $0.1615(5)$               \\
$A(6)_3$  & $6.9463(13)$  &  $6.80(7)$       & $0.1615(4)$               \\\bottomrule
\end{tabular}
\caption{Results for $t_0$, $t^2_0 \chitop$ and its fourth root.
\label{tab:chir2}}
\end{table}

In order to extrapolate the raw data to the continuum and the $N\to\infty$ limit,
we use the functional form
\begin{equation}
t_0^2 \chitop(1/N,a) = t_0^2 \chitop(0,0) +c_1 \frac{1}{N^2} + c_2 \frac{a^2}{t_0} \,,
\label{e:gf}
\end{equation}
which takes into account the leading corrections dictated by the Symanzik and the large-$N$ expansion.
This is  motivated by the observation that both corrections are small, given the statistical accuracy
of our data. The fact that the $N$-dependence of the $\mathrm{O}(a^2)$ term can be neglected within
our precision is further supported by the observation that discretization effects in the ratio
$\chitop^t/\chitop^{t_0}$, which can be captured to exceedingly high accuracy, turn out to be
independent of $N$.

Our main result is obtained by fitting \Eq{e:gf} to the two finer points of \su{4}, \su{5} and
\su{6} data
together with the two finer data points for \su{3},  where the latter is only used to constrain the
coefficient $c_2$.
Discarding the coarser lattice points and the smallest $N$ reduces the assumptions made on the scaling
region of our results. This fit  renders
\begin{equation}
t_0^2 \chitop(0,0)=7.03(13)\cdot 10^{-4}  \,,
\end{equation}
i.e. a 2\% accuracy is reached. The fit quality is excellent
with a $\chi^2/\text{dof}=0.94$. In the continuum limit the fit gives
$t_0^2\chitop(1/N,0)=6.68(12)\cdot 10^{-4}$, $6.81(11)\cdot 10^{-4}$  and $6.87(11)\cdot 10^{-4}$           
for $N=4$, $5$ and $6$ respectively, see right plot of Fig.~\ref{fig:1}.

To get a better handle on possible systematic effects of this result, many
other fits to the data have been tried, all of them leading to similar results.
Among them the most obvious modification is to include also the third finest
point of the \su{3} data determining the discretization effects. This changes the result to
$t_0^2 \chitop(0,0)=7.13(10)\cdot 10^{-4}$ with $\chi^2/\text{dof}=1.1$, compatible with the above number.
If the three finest \su{3} points are globally fitted with the two finer points of the other
groups, the results is $t_0^2 \chitop(0,0)=7.09(7)\cdot 10^{-4}$ with an excellent value of
$\chi^2/\text{dof}=1.0$. A global fit of \Eq{e:gf} to all data, including the three finer
\su{3} ones, adding an $a^2/N^2$ term to \Eq{e:gf}, gives $t_0^2 \chitop(0,0)=7.02(13)\cdot 10^{-4}$ with a
$\chi^2/\text{dof}=1.7$. Performing the continuum limit group-by-group and applying the large-$N$
extrapolation only in a second step also gives a compatible result.

From these analyses we conclude that the systematic effects coming from the continuum and large-$N$
extrapolations are under control within the errors quoted. 

\section{Conclusions}
This is the first investigation of the large-$N$ behavior of the 
topological susceptibility in pure Yang-Mills theory using
a theoretically sound definition of $\chitop$, and small 
lattice spacings which allow for control over the continuum limit.
As a final result we quote for $N\rightarrow\infty$
\begin{equation}
t_0^2 \chitop(0,0)= 7.03(13) \cdot 10^{-4}.
\end{equation}
This result proves that the leading anomalous contribution to the $\eta'$ mass
is large enough to solve the $\mathrm{U}(1)_\mathrm{A}$ problem in QCD. The 
bulk of the mass of the pseudoscalar singlet meson is generated by the anomaly
through the Witten--Veneziano mechanism. The $1/N^2$ corrections that we
have found in $t_0^2 \chitop(0,0)$ are at most of the expected size (even a bit smaller),
with no large prefactor in the expansion.
This explains why the $N=3$ result, $t_0^2 \chitop=6.67(7) \cdot 10^{-4}$,
in \Ref{Ce:2015qha} is already large enough to explain the large value of
the $\eta'$ mass in Nature. The difference with the $N\rightarrow\infty$ value
is barely visible within errors, despite their high accuracy.

In the Yang--Mills theory, it will be challenging to improve significantly on
these results by brute force. Discretization effects and large-$N$ effects are
roughly of the same level. The much higher accuracy needed to resolve higher
order effects in the large-$N$ expansions will  therefore require significantly
smaller lattice spacings. These are still computationally very expensive, even
with the open boundary conditions, which make those used in the present study
possible.

The accuracy presented here is certainly sufficient for the completion of the
proof of the Witten-Veneziano relation in \Eq{eq:wv}. It will need to be
matched by the one on the hadronic quantities entering
the relation to be computed in the large-$N$ limit of QCD.\\

\begin{acknowledgement}
It is a pleasure to thank Rainer Sommer for interesting and helpful discussions.
Simulations have been performed at Fermi and Galileo at CINECA 
(YMlargeN Iscra B project and CINECA-INFN agreement), at the ZIB computer center (project bep00053) with the
computer resources granted by The North-German Supercomputing Alliance (HLRN),
on PAX at DESY (Zeuthen), and on Wilson at Milano--Bicocca. 
We thank these institutions for the computer resources and the technical 
support. M.G.V acknowledges the support from the Research Training Group GRK1504/2
``Mass, Spectrum, Symmetry'' founded by the German Research
Foundation (DFG). 
\end{acknowledgement}

\end{document}